\documentclass[]{spie}  

\usepackage{amsmath,amsfonts,amssymb}
\usepackage{graphicx}
\usepackage[colorlinks=true, allcolors=blue]{hyperref}
\usepackage{subcaption} 
\usepackage{gensymb}

\title{Commissioning of the laser-driven ion acceleration beamline at the Centre for Advanced Laser Applications}

\author[a]{Jens Hartmann}
\author[a]{Thomas F. Rösch}
\author[a]{Felix Balling}
\author[a]{Marc Berndl}
\author[a]{Leonard Doyle}
\author[a]{Lotta Flaig}
\author[a]{Sonja Gerlach}
\author[a]{Luisa Tischendorf}
\author[a]{Jörg Schreiber}
\affil[a]{Department of Medical Physics, Faculty of Physics, Ludwig-Maximilians-Universität München, Am Coulombwall 1, Garching b. München, Germany}

\authorinfo{Further author information: (Send correspondence to J. Hartmann)\\J. Hartmann.: E-mail: jens.hartmann@physik.uni-muenchen.de, Telephone: +49 (0)89 289 14172\\  T.F.Rösch: E-mail: t.roesch@physik.uni-muenchen.de, Telephone: +49 (0)89 289 54023}

\begin{document} 
\maketitle

\begin{abstract}
The Centre for Advanced Laser Applications (CALA) in Garching near Munich features the ATLAS 3000 laser system, which can deliver up to 3\,PW within a pulse length of 20\,fs. It is the driver for the Laser-driven ION (LION) beamline, which aims to accelerate protons and carbons for applications. For commissioning, we currently operate with 5\,J on target in 28\,fs. A $20\degree$ off-axis parabolic mirror focuses the 28\,cm diameter laser-beam down to a micrometer-sized spot, where a vacuum-compatible wave-front sensor is used in combination with a deformable mirror for focus optimization. The nano-Foil Target Positioning System (nFTPS) can replace targets with a repetition rate of up to 0.5\,Hz and store up to 19 different target foils. A dipole magnet in a wide-angle spectrometer configuration deflects ions onto a CMOS detector for an online read-out. Commissioning started mid 2019 with regular proton acceleration using nm-thin plastic foils as targets. Since then proton cut-off energies above 20\,MeV have been regularly achieved. The amount of light traveling backwards from the experiment into the laser is constantly monitored and 5\,J on target have been determined as the current limit to prevent damage in the laser. Protons with a kinetic energy of 12\,MeV are stably accelerated with the given laser parameters and are suitable for transport with permanent magnet quadrupoles towards our application platform. We have performed parameter scans varying target thicknesses to optimize for highest and most stable proton numbers at 12\,MeV kinetic energy, and investigated shot-to-shot particle number stability for the best parameters.

\end{abstract}


\section{INTRODUCTION}
\label{sec:intro}  
Ion sources based on the accelerations in plasmas with high intensity lasers exhibit both promising and at the same time challenging properties in terms of particle spectrum, pulse duration and repetition rate\cite{daido2012review}. Advances in laser-driven ion acceleration research depend to a significant amount on the availability of more powerful laser systems. The ATLAS 3000 laser in the newly build Centre for Advanced Laser Applications (CALA) in Garching near Munich is a Ti:Sa chirped pulse amplification system that has shown to produce pulses with 90\,J before compression and 22\,fs long attenuated compressed pulses. Prospectively, it will be able to deliver pulses of 3\,PW peak power with 1\,Hz repetition rate in regular operation. ATLAS 3000 is the driver for the Laser-driven ION (LION) acceleration experiment in CALA. Its high peak power in combination with its high repetition rate allow the development of a laser-driven ion source suited to a series of applications that demand high particle energies and a large number of particle bunches in reasonable time. One key for these applications is source stability in terms of particle numbers and energies. Ions of certain energies must be available reliably on demand in every accelerated bunch. 
We set up the LION experiment in CALA to be a high repetition rate laser driven ion acceleration experiment with a large degree of flexibility and automation in terms of laser diagnostics and optimization, target positioning, ion transport and online diagnostics. We have been concentrating on creating an integrated system that allows remote setup control and online switching between operation modes in order to reduce time between shots and to exploit the full potential of 1\,Hz repetition rate soon. The near term aim is to deliver on demand 12\,MeV protons in a millimeter size spot to an on air platform which is located at 1.8\,m from the laser target. Here, we present commissioning studies with different experiment parameters and the resulting currently optimized performance of the CALA LION setup and show our setups availability as source for applications and detector tests with laser accelerated protons. 

\section{EXPERIMENTAL SETUP}

\subsection{The ATLAS 3000 Laser and Laser delivery}
\label{sec:laser}
In 2016, after the completion of the CALA building, the setup assembling started in the LION experimental area (the LION cave). In parallel, the infrastructure in the other areas of CALA advanced. Most importantly, the ATLAS laser system was rebuilt and upgraded with two new multipass amplifiers that increased pulse energy to up to 90\,J before compression with up to 1\,Hz repetition rate. However, in order to  reduce damage risks, the applications of the laser for particle acceleration experiments started with strongly reduced laser energies of around 6\,J before compression and are increasing step-by-step with implementation of additional precaution measures. After compression the beam has a flattop beam profile with 28\,cm diameter. The compression is optimized with a feedback loop between pulse duration measurement and an acousto-optic modulation of the stretched pulse. Two detrimental pre-pulses could be eliminated, resulting in a  contrast of $10^{-10}$ compared to ASE level using the partly amplified pulse and measured with a third order autocorrelator.\\
The laser beam delivery (LBD) vacuum system which connects the vacuum laser compressor with the experimental areas was completed in 2018 and successively equipped with mirrors and control systems. Key element of the LBD is a full beam diameter deformable mirror (DM). It features 53 actuators and is part of the post-compressor adaptive optics feedback loop, that is coupled to the laser focus diagnostic in the vacuum chamber. \\
After receiving the radiation and laser safety approval for operation first successful acceleration of protons in the LION-cave was demonstrated on June 26th, 2019.

\subsection{The Laser-driven Ion (LION) Experiment}
\label{sec:setup}

   \begin{figure} [ht]
   \begin{center}
   \begin{tabular}{c} 
   \includegraphics[height=7cm]{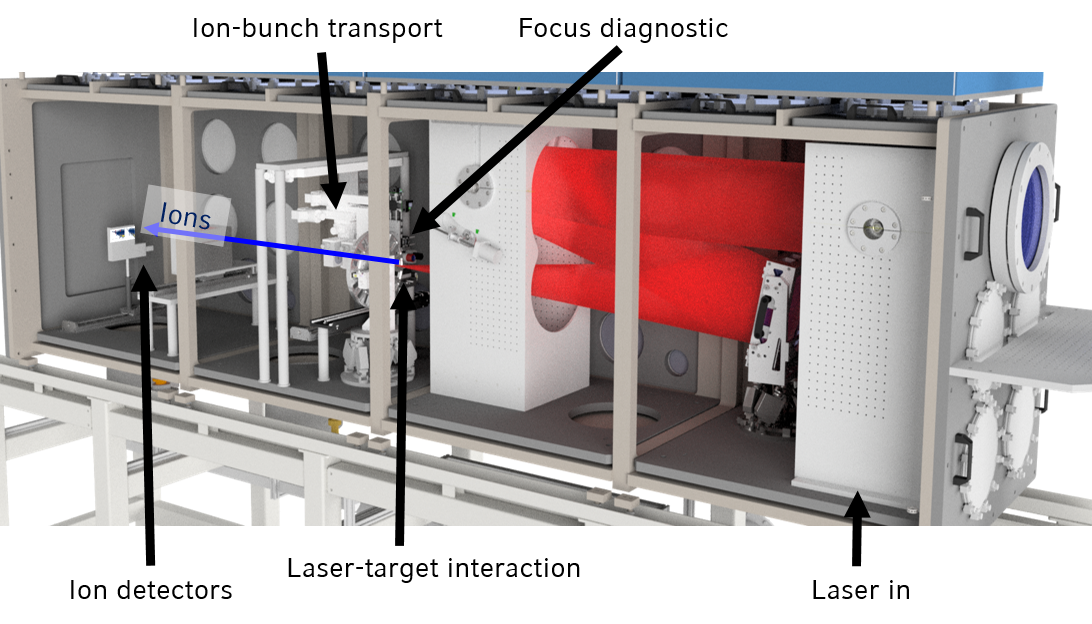}
   \end{tabular}
   \end{center}
   \caption[example] 
   { \label{fig:01_lion_fig2} 
Rendered CAD model of the experimental vacuum chamber and main components for commissioning phase in LION.}
   \end{figure} 

Four nearly cubic modules compose the LION experimental chamber which in total has a length of 4\,m. Fig.\ref{fig:01_lion_fig2} shows the main components of the setup. The laser enters through a port in the bottom of the first module, hits a first dielectric turning mirror with $99\%$ reflectivity in the wavelength region 750 to 850\,nm, and then propagates in the upper half of the chamber onto the second silver-coated turning mirror. This sends the beam onto the silver-coated off-axis parabolic (OAP) mirror in the lower plane of in the first module. It is positioned off-center, focusing the beam under an angle of $6.8\degree$ with respect to the central axis. The effective focal length is 150\,cm so that a flat top beam with 28 cm diameter would be focused to a diffraction limited full-width at half-maximum (FWHM) of 4.6\,$\mu$m. 
The focal plane is simultaneously imaged with $10\times$ "high"-magnification and $3\times$ "low"-magnification onto two CCD-cameras that are integrated in a vacuum compatible microscope. This microscope also incorporates a third imaging path that includes a wavefront-sensor. A loop feeding back from this wavefront-sensor to the DM in the LBD is used to maximize peak intensity in the laser focus.
The microscope assembly also serves for target positioning in combination with a motorized target wheel. This nano-Foil Target Positioning System (nFTPS) has been published by Gao et al\cite{Gao_FTPS_2017}. Since this publication, the nFTPS has been adapted. For example, the aluminum wheel was replaced by a plastic wheel for weight reduction and electromagnetic pulse suppression. It can store up to 19 different target-holders, each containing 40 target-holes. The nFTPS can automatically measure and store the positions of every target hole prior to the experiment. These positions can later be recalled with a repetition rate of up to 0.5\,Hz. The target scanning procedure is as following. First, a crude alignment of the microscope-target distance is performed with a confocal distance sensor, which is mounted parallel to the microscope objective. Then, the offset of the target-hole centers to the laser-focus center is measured from images taken by the microscope and stored for each hole. Each target-hole distance is then measured and corrected again based on the previous measurement. After all target positions are stored within the software program, a second routine can be started that takes images of all target foils for later reference. These also can be previewed during target selection. Finally, before performing the laser experiments, the microscope is moved out of the beam path into a safe position.\\
Sending a high power laser-pulse into the target chamber is referred to as a laser-shot. Prior to the laser-shot a target has to be selected from the target wheel program. The laser-shot event can then be triggered either manually (on demand) or automatically once the system has stopped moving. 200\,ms prior to the laser-pulse being generated, a trigger pulse is sent to the experiment to prepare diagnostics such as cameras and ion-detectors to record during the laser-plasma interaction. The data is then stored automatically for each armed diagnostic in a separate folder on a networked drive. All data regarding the same laser-shot can be identified by having a distinct shot-number in their filename.\\
A pair of permanent magnet quadrupoles can be driven into the ion-bunch path to transport and focus the laser-accelerated protons. Each quadrupole consists of 12 NdFeB wedges magnetized according to the Halbach design\cite{Halbach19801}. The quadrupoles are 40\,mm and 20\,mm long and have a 1\,cm diameter bore with around 330\,T/m magnetic field gradient. Both magnets are mounted on a motorized positioning stage with two motorized axes to position the quadrupoles along the beamline. A third motorized axis additionally allows to insert or remove the quadrupoles from the beamline in transverse direction. The ion drift lengths from the target to the first quadrupole and between two quadrupoles are chosen such that protons of a selected design energy are focused on a LANEX$^{TM}$ scintillating screen that is located 2.5\,cm behind a vacuum exit window on air. For protection of laser stray light this 50\,$\mu$m thin kapton window is covered with a 15\,$\mu$m aluminum sheet. Because of geometric drift length limitations the minimum design energy for protons is 12\,MeV. For protection the first quadrupole is equipped with an elliptical aperture that limits the incoming bunch divergence in a way that proton absorption in the internal walls of the quadrupoles is minimized. Further, in order to protect the laser and the targets from reflected light and debris from the quadrupole front side, a glass shielding tube is inserted between targets and the first quadrupole.

\begin{figure}
  \centering
    \includegraphics[width=\textwidth]{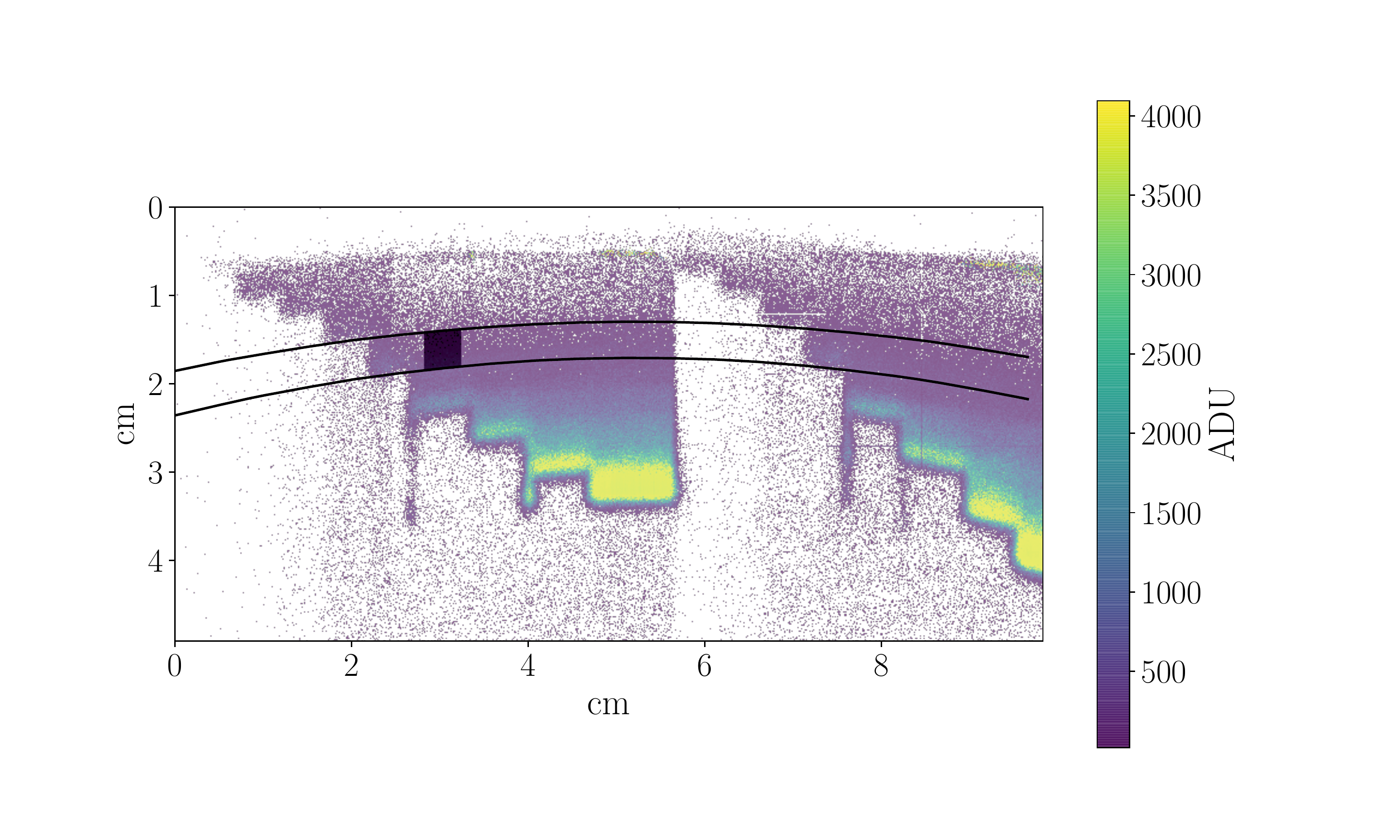}
    \caption{Typical background subtracted image from the Radeye1 detector. The horizontal bows are equi-energy lines, with 8.8\,MeV and 12.2\,MeV. For the evaluation, pixels within the marked area have been summed up.}
    \label{fig:radeyedata}
\end{figure}

As primary ion diagnostic a wide-angle spectrometer (WASP) is used \cite{lindner_novel_2018}. A slotted aluminum plate with a slit width of 200\,$\mu$m is positioned 82\,cm downstream of the target. It cuts the particle spray emitted in this direction in a horizontal fan beam which thereafter enters a magnetic dipole field with a gap size of 10.5 cm. The field is created by permanent magnets in an iron yoke and is therefore inhomogeneous with a minimum magnetic field strength of order 165\,mT in the center. The positively charged ions are deflected downwards and captured on a CMOS sensor (Radeye1) with 48\,$\mu$m pixel size that covers an area of 10\,cm by 5\,cm. The CMOS sensor is covered with an aluminum template whose thickness varies between\,100 $\mu$m and 2.2\,mm in 8 discrete steps. This yields distinct kinetic energy values given by the stopping of the ions in aluminum of the respective thickness. Fig.\ref{fig:radeyedata} shows one example image of this CMOS signal.
For a first qualitative information on the source performance, it is sufficient to sum over all pixel values, called Analog to Digital Units (ADU), on a background subtracted image in an area of interest. 
The detector is sensitive to single protons and reacts linear to deposited energy, as studied by Reinhard et al.\cite{reinhardt2011test}.
Based on the visual information provided by the aluminum phantom we have derived the following method:
The aluminum stripes with thicknesses 280\,$\mu$m, 500\,$\mu$m, 880\,$\mu$m and 1460\,$\mu$m require a minimum kinetic proton energy of 6.3\,MeV, 8.8\,MeV, 12.2\,MeV and 16.3\,MeV for protons to be transmitted towards the active detector surface. Two stripes for each thickness are mounted on the left and right side respectively. Between these known points we draw equi-energy lines on our raw-image, as seen in Fig.\ref{fig:radeyedata}. Our interest in the system stability with our beamline at 12\,MeV resulted in evaluating the interval between 8.8\,MeV and 12.2\,MeV, which has a center of 10.5\,MeV. This determines the horizontal borders of our area of interest. The vertical borders where chosen to be within the aluminum strip corresponding to 6.3\,MeV transmitted energy. A more sophisticated evaluation method is currently prepared for publication.
The dipole magnet and the detector are both mounted on motorized stages and can be removed from the ion-bunch path.

\section{PERFORMANCE AND COMMISSIONING STUDIES}
\begin{figure}
  \centering
  \begin{subfigure}[b]{0.4\textwidth}
    \includegraphics[width=\textwidth]{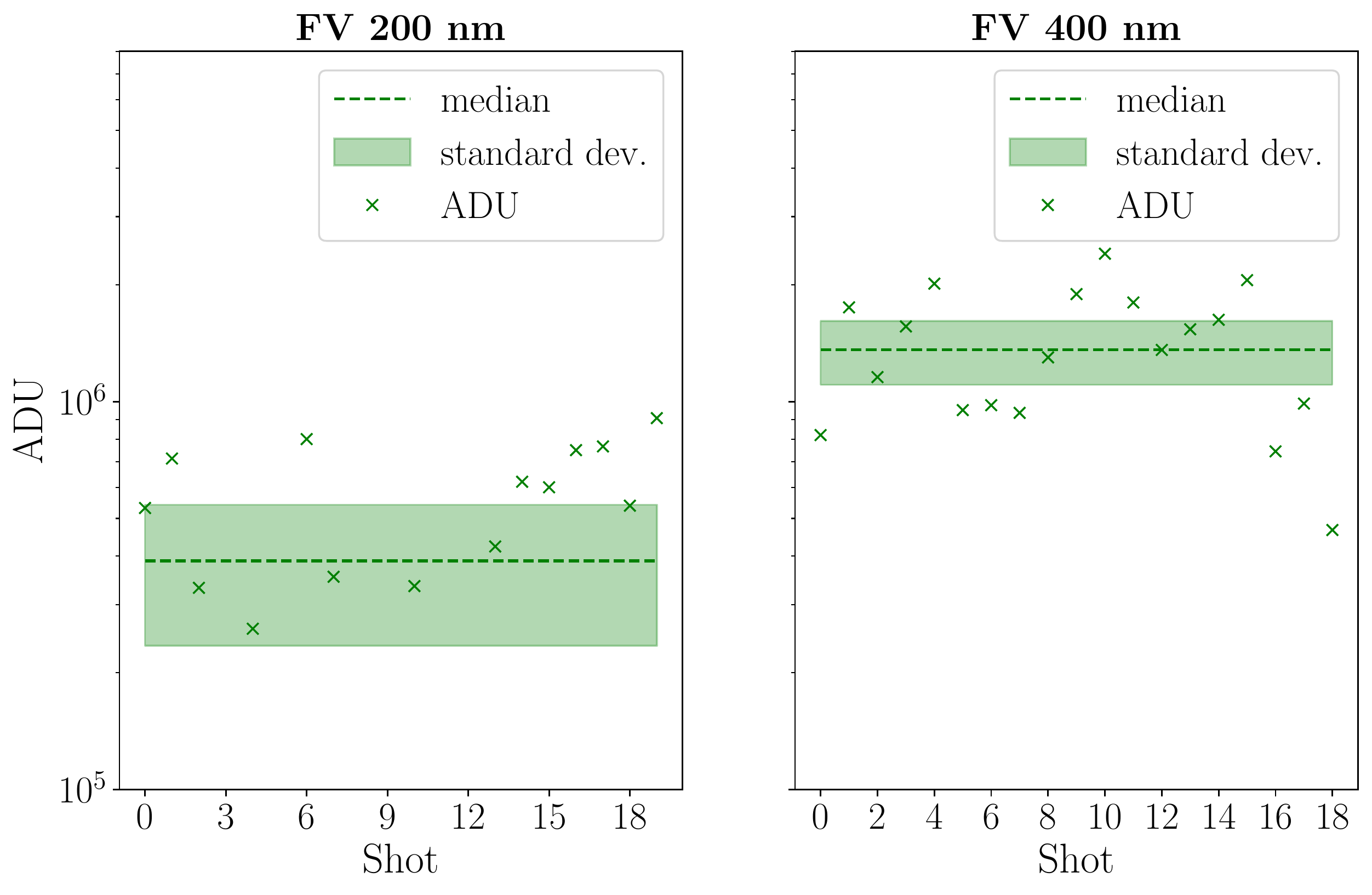}
    \caption{Performance of 200\,nm and 400\,nm formvar (FV) target foils on the same day with the same laser parameters, evaluated at 10.5$\pm$1.7\,MeV. The summed ADU inside the evaluated bin correlate with the incident particle number. 
    }
    \label{fig:particlestability}
  \end{subfigure}
  \begin{subfigure}[b]{0.4\textwidth}
    \includegraphics[width=\textwidth]{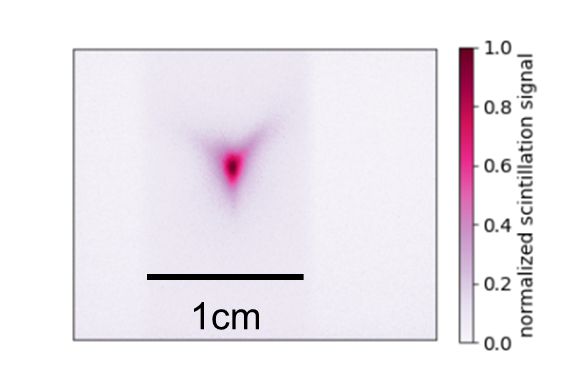}
    \caption{Smallest proton focus achieved on the scintillating screen from 400\,nm formvar foils with quadrupole doublet set to 12\,MeV design energy and 1.8\,m focus distance.\\}
    \label{fig:protonfocus}
  \end{subfigure}
\end{figure}

We started operation and commissioning of the setup in 2019 with around 2\,J laser energy on target and first protons with kinetic energies up to 10\,MeV were observed in June 2019. 
Later that year we achieved the first proton focus with the permanent magnet quadrupole doublet on November 30th, 2019 with 12\,MeV design energy. Since then, we worked on stabilizing and improving the performance in terms of maximum energy and stability.\\
One step was to further increase the laser energy on target from 2\,J in 2019 to 5\,J by the end of 2020. To monitor whether the light scattered back from the laser-target interaction was reaching damaging levels inside the laser, one photo-diode was installed in the front-end and one in the final amplifier of the ATLAS 3000. Their positions allow them to detect out-going and returning light. Based on damage estimates and a cross-calibration of the front-end diode a detection threshold was defined for the signal. As long as the detected light was below this threshold, the front-end was determined to be save. This threshold was almost reached with about 5\,J laser-pulse energy on target, which we achieved during the second half of 2020 with a laser-pulse length of 28\,fs and a laser-focus radius of 2.46\,$\mu$m with 25\% enclosed energy.
We also measured the laser contrast on a regular basis and observed a very stable contrast over many months. The contrast together with the pulse energy determines mostly the optimum target thickness for best acceleration performance. We performed a series of commissioning experiments to find the best working target under the given laser conditions. We tested plastic (formvar) and gold foils as target materials. For the same target thickness we found little difference in performance between plastic and gold. Plastic however had the advantage of reduced debris and fratricide and it was easier to manufacture with different thicknesses. We investigated formvar thicknesses ranging from 20\,nm to 800\,nm and found the most reliable performance with targets of 400\,nm thickness. A comparison of particle numbers between multiple shots for 200\,nm and 400\,nm is shown in Fig.\ref{fig:particlestability}.\\
The laser energy increase together with the adaptive optics loop assisted focus optimization and the improvement in target selection and operation resulted in cutoff energies in the proton spectrum above 20\,MeV and the acceleration of protons to energies beyond 15\,MeV was reliably possible.

This reliable operation enabled first detailed studies of proton beam guiding with the permanent magnet quadrupole doublet. Due to the large number of shots that is now possible, we could fine tune the quadrupole alignment online on the proton beam and minimize both transverse beam steering and focus shape. We could reduce the transverse proton focus size in the focal plane to about 1-2\,mm (Fig.\ref{fig:protonfocus}), being an acceptable size for a wide range of applications and detector tests. The increased fluence in the optimized proton focus also allowed further investigation of the maximum cutoff energies. Due to the energy selective behaviour of a permanent magnet quadrupole doublet\cite{rosch2017considerations} in the focus we could gradually increase the design energy and observe at which energy the scintillating signal vanished over multiple shots. This experiment confirmed the 20 MeV cutoff energy. In another experiment instead of the scintillating screen we placed a stack of radiochromic EBT3 films in the focus and set the design energy of the doublet to 20 and 23\,MeV, respectively. At both configurations we integrated 20 laser shots on the stack. The evaluation of the proton stopping in the stack confirmed the maximum energy of 20\,MeV and even a signal at 23\,MeV could be observed. These studies reproduced and confirmed the performance already observed with the WASP and yielded further certainty about the general setup performance.\\
We recorded data on the light transmitted through the laser-target interaction as well as scattered light from the illuminated side of the target. This gives a first non invasive feedback on the proton acceleration and is a very useful tool. For example, complete transmission evidences the lack of a target. Further, correlating these light-based signals and their corresponding proton spectra might allow to quantify this information for real non invasive acceleration monitoring for applications.

\section{CONCLUSIONS}
We can regularly operate LION at CALA with laser pulses of $\approx 200$\,TW peak power and intensities beyond $10^{20}\mathrm{W/cm^2}$. The temporal contrast at this level limits applicable (plastic) foil thicknesses to 200 to 400\,nm. These values result in proton energies beyond 20\,MeV and allows generating a proton focus with 12\,MeV on a regular basis. We are now in good position for first irradiation studies which can accept shot-to-shot fluctuations if monitored accordingly. To foster such studies, we currently implement reliable and quantitative proton bunch instrumentation in the proton beamline behind the quadrupole doublet and in the proton focus. With those in place, we can start detailed studies to further optimise the proton focus and the related deposited energy density distribution. Such studies will foster understanding the origin of remaining fluctuations and are prerequisite to provide feedback for active stabilisation methods. 

\acknowledgments 
This work was supported by the BMBF under project 05P18WMFA1. SG acknowledges the support of the German Research Foundation (DFG) within the Research Training Group GRK 2274. LD acknowledges the support of the DFG within the project FOR 2783/1. We thank the complete working group of Prof. Stefan Karsch from the LMU for active support with the laser development, monitoring and operation. Further, we thank the CALA team around Hans-Friedrich Wirth, especially Jerzy Szerypo for our target production and Oliver Gosau and Nikolce Gjotev for setting up and maintaining the LBD.
\bibliography{report} 
\bibliographystyle{spiebib} 

\end{document}